\documentclass[sigconf,nonacm,10pt,natbib=false]{acmart}
\usepackage[english]{babel}
\usepackage{blindtext}
\usepackage{multirow}
\usepackage{algorithm}
\usepackage{url}
\usepackage{algpseudocode}

\long\def\comment#1{}
\long\def\delete#1{}

\newcommand{\ie}{{i.e.}}
\newcommand{\eg}{{e.g.}}

\newcommand{\highlight}[1]{{ #1}}

\newcommand{\newadd}[1]{{ #1}}

\AtBeginDocument{%
  }

\setcopyright{acmcopyright}
\copyrightyear{2018}
\acmYear{2018}
\acmDOI{XXXXXXX.XXXXXXX}

\RequirePackage[
  datamodel=acmdatamodel,
  style=acmnumeric,
  ]{biblatex}

\title{Optimize Replica Server Placement in a Satellite Network}

\author{\large Zhiyuan He$^1$, Yi Xu$^{2}$, Cheng Luo$^{1}$, Lili Qiu$^{1}$, Yuqing Yang$^{1}$}

\affiliation{
  \vspace{5pt}
  \institution{$^1$Microsoft Research, $^2$USTC}
  \country{}
  \vspace{15pt}
}
\renewcommand{\shortauthors}{%
Z. He, Y. Xu, C. Luo, L. Qiu, Y. Yang
}

\begin{document}

\begin{abstract}

Satellite communication offers Internet connectivity to remote locations, such as villages, deserts, mountains, and at sea. However, transmitting content over satellite networks is significantly more expensive than traditional Internet. To address this issue, we propose placing content replica servers within satellite networks and optimizing replica placement for important performance metrics, such as latency, transmission, and storage cost. Our approach can support different types of satellite networks, including Low Earth Orbit (LEO), Medium Earth Orbit (MEO), Geostationary Orbit (GEO), and their combinations. An important challenge for supporting content replicas in such networks is that LEO and MEO satellites are constantly moving. We address this challenge by explicitly considering their moving trajectories and strategically optimizing not only client performance, but also the cost of transferring content from one satellite to another as needed. We demonstrate the effectiveness of our approach using both simulated traffic traces and a prototype system.

\end{abstract}

\maketitle

\section{Introduction}
\label{sec:intro}

The first satellite was launched in 1957 \cite{divine1993sputnik}. Since then satellite communication has gone through rapid development. We now have Geostationary Orbit (GEO), Medium Earth Orbit (MEO) and Low Earth Orbit (LEO) satellites. GEO, MEO, and LEO are 35,785 km, 2,000 -- 36,000 km \cite{su2019broadband}, and less than 2,000 km \cite{qu2017leo} from the Earth's surface, respectively. 

GEO satellites are stationary, simplifying communication with ground stations. In comparison, both MEO and LEO satellites appear in motion to the ground station, and require the ground station to steer in different directions to stay connected and need to switch to a different satellite when the previous one goes out of sight \cite{park2021trends}. Moreover, they require more satellites to cover the Earth due to their closer distance to the Earth. In return for the larger number of satellites and management complexity, MEO and LEO satellite communication enjoys lower latency and higher bandwidth owing to their proximity to the Earth. LEO satellite communication features the lowest delay and highest bandwidth, and can potentially support challenging real-time applications \cite{michel2022first}. In this paper, we consider all types of satellites but focus most on LEO networks due to its support of real-time applications and increasing commercial interest. For example, Starlink has over 2,600 LEO satellites \cite{clark_2022}, while Amazon plans to launch over 3000 LEO satellites in the next few years \cite{amazon}.




\begin{figure}[tp]
\centering
\includegraphics[width=0.45\textwidth]{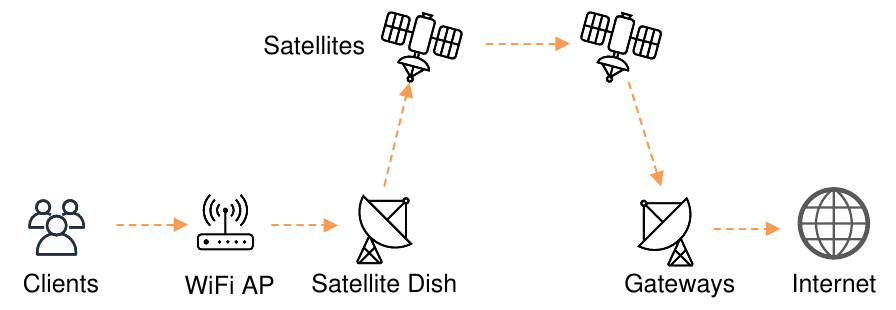}
\caption{A typical path between a client and the Internet in LEO satellite network.}
\label{figure:user-satellite-typical-link}
\end{figure}

As shown in Figure~\ref{figure:user-satellite-typical-link}, a client using a LEO satellite communicates with an Internet server through the following path: its device $\xrightarrow{}$ its WiFi router $\xrightarrow{}$ its satellite dish $\xrightarrow{}$ a satellite $\xrightarrow{}$ 0 or more satellites $\xrightarrow{}$ a gateway closest to the server $\xrightarrow{}$ the server. This path involves two hops between ground and satellites and 0 or more hops among satellites. The delay can be significant.


\begin{figure}[tp]
\centering
\includegraphics[width=0.45\textwidth]{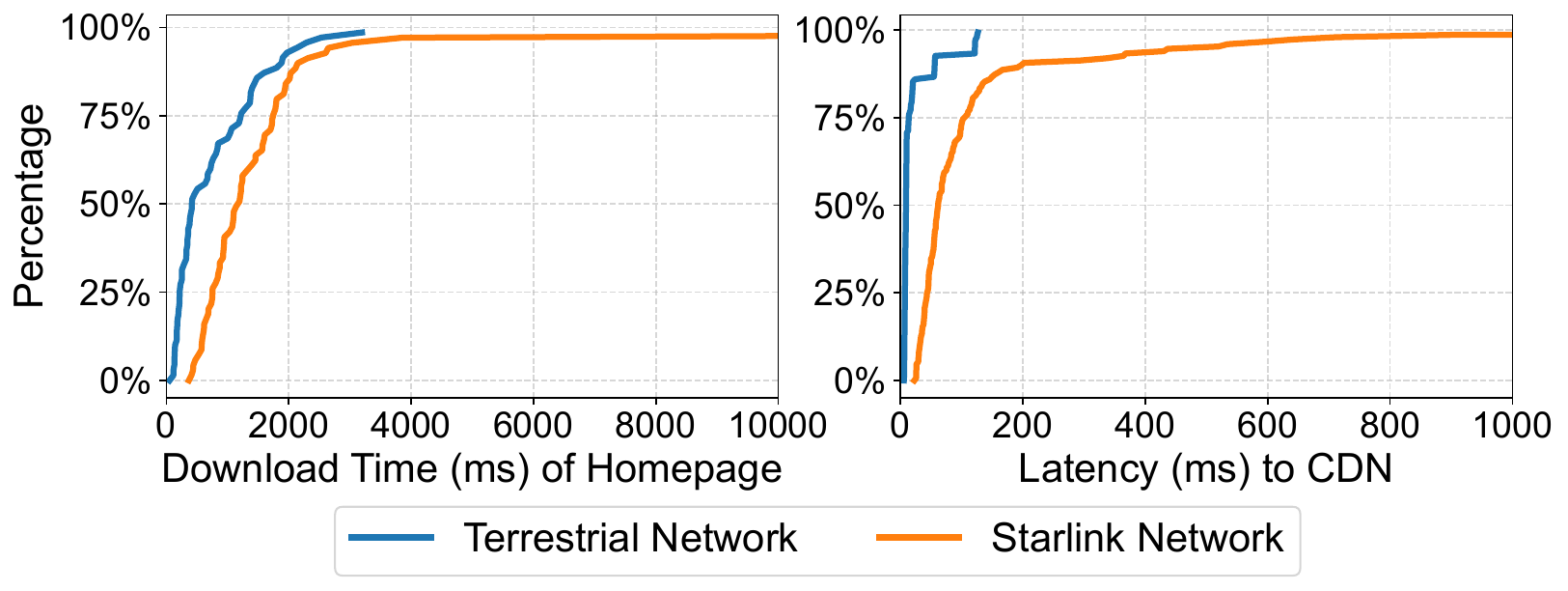}
\caption{Measured download time of web homepage and latency to CDN in satellite network and terrestrial network.}
\label{figure:testbed-measured-time}
\end{figure}

We measure Web performance using a Starlink ground station in Texas. Download time of the top 100 websites' homepages \cite{pochat2018tranco} and latency to the top 15 content delivery network (CDN) providers \cite{cdnperf} are compared to the terrestrial network at the same location.  Figure \ref{figure:testbed-measured-time} shows that the median time to download a website homepage in Starlink is 2.7 times of terrestrial network, with a median latency to a CDN server 7.1 times. In addition, we observe a long tail in both the webpage download time and CDN latency in the Starlink satellite network likely due to satellite movement. 



The significant end-to-end delay in LEO satellites motivates our design of CDN for satellite systems. In the existing Internet, CDN is widely used to speed up content delivery and reduce delivery cost \cite{barnett2018cisco}. \highlight{
Advancements in hardware and space technology have made it plausible to place storage and computing devices in the sky. According to \cite{bhattacherjee2020orbit}, a modern server constitutes only 6\% of a Starlink satellite weight and consumes only 15\% of its harvested solar energy. Although these costs are notable, they are not prohibitive. The in-orbit computing device can  substantially decrease the latency experienced by satellite users and reduce the bandwidth requirements.}

Different from the traditional CDN, the relative position of LEO satellites and ground stations are constantly changing, so we can no longer place content on the same satellite to serve the same region. For example, as the satellite serving the region is moving away, its content may need to be replicated to another satellite in order to serve the same region. In general, we should explicitly take into account of the LEO orbits in order to design the CDN for LEO satellite communication. 

More specifically, we formulate the replica placement problems for satellites as an optimization problem that selects servers to minimize the total cost of (i) delivering the content from replicas to users, (ii) replicating the content to replica servers, and (iii) the storage cost for the content. 

We develop efficient algorithms that leverages satellite orbits, and evaluate our approaches through both simulation and emulation, in both web browsing and video streaming scenarios. The simulation experiments are based on three real web traces and use real latency values from the Starlink network. In addition, we implement a real-world prototype system that employs separate processes to mimic clients and servers, and uses data from the Starlink network to emulate network connections. Our results show that our methods can not only significantly enhance user experience, but also reduce replication cost and storage costs. We will release our code to the community. 
\section{Background}
\label{sec:background}


\subsection{Content Distribution Network (CDN)}

CDN replicates content, such as web pages, software, audio, and video, across a distributed set of nodes so that a client can fetch the content from a nearby CDN node, thereby offloading the origin server, improving the latency, and reducing the cost. Some of the popular CDN providers include Akamai, Amazon AWS, Microsoft Azure, Google cloud. 

On the research front, a number of theoretical models have been proposed for replica server placement \cite{sahoo2016survey}, including facility location (\ie, minimizing the total cost of opening facilities and the cost of using the facilities to serve the clients)~\cite{sung2013efficient}, K-median (\ie, selecting a given number of replicas to minimize the total cost of serving the clients)~\cite{placement01} and K-center (\ie, similar to K-median but minimizing the maximum cost instead of the total cost)~\cite{jin2016toward}. A series of algorithms have been developed to tackle the problems and most of them use greedy or other heuristic algorithms \cite{sahoo2016survey}. 

Current CDNs are designed for stationary networks. However, the network topology is constantly changing in satellite networks. If we place content on satellites, the satellites will needs to replicate it to other nodes when they move away from the interested regions. We need to explicitly consider replication costs caused by satellite movements when designing a CDN for satellite networks. 


\subsection{Satellite Networks}

Artificial satellites are launched into the space to support a variety of applications, such as communication \cite{huang2020recent}, weather monitoring \cite{rao1990weather}, and navigation \cite{grewal2007global}. GEO satellites are the earliest communication satellites \cite{evans20111945}. Since they have the same orbital period as the Earth, these satellites appear stationary to the ground station so that the ground station can steer in the same direction for communication. However, their large distance from the earth (35,785 km) makes it impossible to support time-critical traffic. To reduce the latency, LEO satellite communication is becoming popular \cite{L2D2}. Starlink has launched over 2,600 LEO satellites \cite{clark_2022}, OneWeb has over 600 satellites in orbit \cite{oneweb600}, and Amazon plans to launch 3,236 satellites in the next few years~\cite{amazon}. LEO satellites orbit around the Earth much faster than the Earth's orbital period. For example, the orbital period of a StarLink satellite is 95 minutes. MEO satellites are in between GEO and LEO in terms of their altitude, moving speed, and a number of satellites required to cover the Earth. 

Satellite networks have high communication costs, making CDNs appealing for them. Unlike traditional Internet, where nodes are typically stationary, LEO and MEO satellites are constantly moving with respect to the end users and we need to dynamically move the content around to satisfy the same region. This motivates us to develop replication algorithms for satellite networks.
\section{Our Approach}


\subsection{Problem Formulation}
\label{sec:problem-formulation}

Our goal is to develop algorithms to determine the replica placement that minimizes the total cost of replicating the content and delivering it to end users during a given period. 

First, to formulate the problem, we use a time-dependent graph to fully reflect the dynamics of satellite network. We partition time into discrete slots $t$ and represent the network as a time-dependent graph $G_t = <V, E_t>$. $V$ consists of three types of vertices:  $V_{user}$ represents users or merged user regions, $V_{replica}$ represents potential replica servers such as satellites, ground stations, and terrestrial servers, and $V_{origin}$ represents the origin server, which always holds the content. The edge set $E_t$ is associated with time $t$ and captures dynamic connectivity between nodes. For example, one satellite $v_a$ can be connected to a ground station $v_b$ at time $t_1$ but disconnected at time $t_2$. In this scenario, we will always have $v_a, v_b$ in the vertex set $V$ but the edge between $v_a$ and $v_b$ will be in $E_{t_1}$ but absent in $E_{t_2}$.

Second, users' queries to content are also dynamic. They generate requests for different content during each time slot. We choose popular files and represent them as a content set $C$. $size_c$ represents the size of file $c \in C$, and $demand_{v, c, t}$ represents the quantified request demand from user $v$ for content $c$ at time slot $t$. 

Our task is to choose suitable replica servers for each content over a given period. We define the replica set for content $c$ in time slot $t$ as $S_{c, t}$. The algorithm should choose $S_{c, t}$ given network topology and user demand. Origin servers are always included in $S_{c, t}$ ($V_{origin} \subset S_{c,t}$). We add additional replica servers $v \in V_{replica}$ to $S_{c, t}$. The potential replica server set $V_{replica}$ consists of satellites, ground stations, and terrestrial servers. The replica set $S_{c, t}$ is time-dependent, allowing it to reflect dynamic transfer and replication of content between satellites. For example, if a satellite $v_a \in S_{c, t}$ moves away from a serving region, it can be replaced in $S_{c, t + 1}$ by another satellite $v_b$. 

In CDNs, selecting the ideal location for replica servers is a key issue. Typically, cost functions are utilized to assess different placement, and we formulate it as an optimization problem \cite{replica-survey}. According to \cite{replica-survey}, the cost objectives usually include (1) Deployment costs for computing, storage, and bandwidth resource, (2) Delivery costs for distributing the content to end users, and (3) Update costs for content revisions. In our study, we ignore the update costs, viewing it as less critical than others. Besides, given the scarcity of bandwidth and storage on satellites, we make a clear distinction between traffic replication costs and storage costs to refine our analysis. We also use the term "query cost" instead of delivery cost in this paper for straightforward comprehension. Note that our algorithms are general and can easily be applied to support other performance metrics.

\noindent \textbf{Query cost}: Query cost quantifies the cost of serving user requests. We assume each user will query the closest replica and the total query cost will be proportional to the user number and demand. Specifically, the query cost is 
$$
\sum_c{\sum_t{\sum_{v_{user} \in V_{user}}{demand_{v_{user},c,t}}\times \min_{v\in{S_{c, t}}}{cost^t_{query}(v_{user}, v)}}}
$$
where $cost^t_{query}(\cdot, \cdot)$ is a cost function between the user and the server vertex. It is derived from the graph $G_t$, and can be quantified by hop count or latency. \highlight{Following \cite{lai2021cooperatively,Handley-hotnets18}, we employ the shortest path algorithm for routing; we always select the nearest replica with the minimum query cost.}

\noindent \textbf{Replication cost}: We use replication cost to represent the costs of creating new replicas. It is formulated as 
$$
\sum_c{\sum_t{\sum_{v_{new} \in S_{c,t}}{\min_{v_{old} \in S_{c, t-1}}{cost^t_{replication}(v_{new}, v_{old})}}}}
$$
where $cost^t_{replication}(\cdot, \cdot)$ is a cost function representing the cost of replicating content from an old replica to a new replica. If $v_{new} = v_{old}$, we will have $cost^t_{replication}(v_{new}, v_{old}) = 0$. A unique aspect of our formulation is that our network is dynamically changing and the replication cost between two nodes may change depending on their relative position at that time. 
This is due to the orbital organization of satellites, where replication to adjacent satellites in the same or neighboring orbits is typically more cost-effective than to distant satellites. We encourage content replication to occur as close as possible to reduce replication traffic. 


\noindent \textbf{Storage cost}: We formulate storage cost as 
$$
\sum_c{\sum_t{\sum_{v \in S_{c,t}}{size_c \times cost_{storage}(v)}}}
$$
where $cost_{storage}(\cdot)$ is a cost function to quantify the cost of storing one unit size of content at different servers. Storing content on satellites is typically more costly than storing at ground servers \cite{lai2021cooperatively}. By assigning different storage costs to different server types, we can effectively utilize various servers for optimal results. 

Note that our definitions of query cost, replication cost, and storage cost are based on general functions. The specific function definition can be modified as needed in practical scenarios. In our research, we focus on common scenarios: (1) Both query cost and replication cost are based on network topology, which means users will typically  query the closest replica server in network, and content replication is preferred between nearby satellites. (2) The storage cost is directly proportional to the content size, with the satellite storage more expensive compared to terrestrial storage. We also study the effects of different storage cost ratios in Section \ref{sec:gateway}. Our formulation is flexible enough to support other cost models and requirements. For example, to enforce a storage constraint, one can set $cost_{storage}(\cdot)$ to 0 up to a certain storage threshold and increase rapidly after exceeding a threshold.

\subsection{Our Replication Algorithms}

Our problem is to find $S_{c,t}$ that minimizes the sum of the three costs (\ie, query cost, replication cost, and storage cost) across all time slots, where $S_{c,t}$ is chosen from $V_{replica}$. If  only one time slot is considered, we can transform our problem to a well-known NP-hard problem -- the uncapacitated facility location problem (UFL) \cite{mirchandani1990discrete}, by treating the sum of replication cost and storage cost as the opening cost of a facility, and query cost as the transport cost in UFL. Since only solving one time slot is already NP-hard, our general multi-time-slot problem is also NP-hard. In this section, we will introduce our practical algorithms. We omit the dimension of content $c$ in this part for ease of description because the algorithm can be applied to each content independently.  

\subsubsection{Multi-time Local Search}
\label{sec:mtls}

We look for an effective algorithm that works well in practice. Inspired by the local search algorithm for UFL \cite{arya2001local}, we first propose a multi-time local search (MTLS) method. The UFL local search algorithm operates as follows: Given an initial facility set $F$, the algorithm tries to find a nearby solution of $F$ with an improvement in the objective, by addition, removal, or replacement of a facility in $F$. The algorithm continues this process until no further improvements can be made. 

Suppose there are $T$ time snapshots. Our task is to find $T$ replica set $S_1, S_2, ..., S_T$. Similarly, we can define the nearby solution of each $S_t$ to be:  $|S_t^{nearby} \setminus S_t| = 1$ or $|S_t \setminus S_t^{nearby} | = 1$. For the sake of efficiency, we do not test $S_t^{nearby}$ one by one. Instead, we solve the following sub-problem: Given $\{S_1, S_2, ..., S_T\}$, what is the best solution among all possiblities of $\{S_1^{nearby}, S_2^{nearby}, ..., S_T^{nearby}\}$. Recall that we always have $V_{origin} \subset S_t$ and we need to choose new replicas from $V_{replica}$. Let $N = |V_{replica}|$. For each $S_t$, the total number of $S^{nearby}_t$ will be $O(N^2)$, where adding a new replica yields $O(N)$ sets, deleting a replica yields $O(N)$ sets, and replacing a replica yields $O(N^2)$ sets. Using an exhaustive search method to acquire the best solution will take $O(N^{2T})$.

We use dynamic programming (DP) to avoid the exhaustive search and enhance efficiency, which takes $O(TN^4)$. To introduce our DP, we first introduce some symbols as follows. We define $f(t, S^{nearby}_t)$ to represent the minimum total cost for time $1,2...t$ and we are required to use replica set $S^{nearby}_t$ at time $t$. $\hat{S}^{nearby}_{t-1}$ is defined as the replica set used at $t - 1$. $RC(t, \hat{S}^{nearby}_{t-1}, S^{nearby}_t)$ is defined as the replication cost from $\hat{S}^{nearby}_{t-1}$ to $S^{nearby}_t$ following our definition in Section \ref{sec:problem-formulation}, while we use $QC(t, S^{nearby}_t)$ and $SC(t, S^{nearby}_t)$ to represent the query cost and storage cost at time $t$ when we choose $S^{nearby}_t$ as the replica set.

We are required to use $S^{nearby}_t$ for $f(t, S^{nearby}_t)$, thus the query cost and storage cost at time $t$ are fixed, represented as $QC(t, S^{nearby}_t)$ and $SC(t, S^{nearby}_t)$, respectively. Besides, to minimize $f(t, S^{nearby}_t)$, we should minimize sum of replication costs at time $t$ ($RC(t, \hat{S}^{nearby}_{t-1}), S^{nearby}_t)$ and total costs before time at time $t-1$ ($f(t-1, \hat{S}^{nearby}_{t-1})$). We enumerate all possible $\hat{S}^{nearby}_{t-1}$ and assign the smallest $RC(t, \hat{S}^{nearby}_{t-1}, S^{nearby}_t) + f(t-1, \hat{S}^{nearby}_{t-1}) + QC(t, S^{nearby}_t) + SC(t, S^{nearby}_t)$ to $f(t, S^{nearby}_t)$. Details are shown in Algorithm \ref{alg:mtls}.

\newadd{Our DP can be proved by mathematical induction. There is no user demand at $t=0$ thus $f(0, V_{origin})=0$ is the best for $t=0$. Now assuming that $f(t-1, \hat{S}^{nearby}_{t-1})$ is the smallest cost for time $1,2,..,t-1$,  $f(t, S^{nearby}_t)$ has to be constructed from $t-1$ and we have enumerated all possibilities of replicas we can use in time $t - 1$. Otherwise, there is a contradiction if we check the replica we use at $t-1$.}

The DP's time complexity $O(TN^4)$ comes from the following calculation. The number of total states in DP is $TN^2$, where $N^2$ is the number of nearby set ($S^{nearby}_{t}$). To get every state, we need to enumerate $N^2$ nearby sets ($S^{nearby}_{t-1}$) at time $t-1$. Thus the DP process will take $O(TN^4)$.

\begin{algorithm}[H]
\caption{Multi-time Local Search (MTLS)}
\label{alg:mtls}
\begin{algorithmic}[1]
\Require Time slot number $T$, Origin server set $V_{origin}$, Replica candidate set ${V_{replica}}$, Query cost function $QC$, Replication cost function $RC$, Storage cost function $SC$, Max iteration number $M$
\State $S_t = V_{origin}, t=1,2,..,T$
\For{$m = 1$ to ${M}$} \Comment{iteratively update $S_t$}
    \State $f(0, V_{origin}) = 0$
    \For{$t = 1$ to $T$}
        \For {each $S^{nearby}_t$ at time $t$}
            \For {each $\hat{S}^{nearby}_{t-1}$ at time $t-1$}
                \State $\hat{c} = QC(t, S^{nearby}_t) + SC(t, S^{nearby}_t) + RC(t, \hat{S}^{nearby}_{t-1}, S^{nearby}_t) + f(t-1, \hat{S}^{nearby}_{t-1})$
                \State $f(t, S^{nearby}_t)=min\{\hat{c}, f(t, S^{nearby}_t)\}$
            \EndFor
        \EndFor
    \EndFor
    \State Update $S_t, t=1,2,..,T$ according to $f$
\EndFor
\Ensure $S_t, t=1,2,..,T$
\end{algorithmic}
\end{algorithm}

The replacing operation yields $O(N^2)$ nearby sets, which results in $O(N^4)$ time complexity. However, if we limit replacements to the $k$ nearest neighbors of a replica in graph $G_t$, the replacing operation will only yield $O(kN)$ candidates and the time complexity will be reduced to $O(Tk^2N^2)$. It makes sense because replicating to nearby satellites incurs lower replication costs. In our evaluation, we set $k=4$. We call this algorithm multi-time local search (MTLS) and its overall time complexity is $O(MTk^2N^2)$, where $M$ is the maximum number of iterations.

\subsubsection{Multi-time Orbit-based Local Search}
The MTLS algorithm is slow in practice due to the large value of $N$ in modern LEO satellite constellations. To overcome this issue, we propose utilizing the satellite orbit information. 


Satellites are typically organized into multiple orbits, for instance, StarLink phase I has 72 orbits with 22 satellites each \cite{starlinkphase1}. Given the fixed trajectory of orbits, we propose a hierarchical method for selecting satellite replicas. First, we select an orbit for each time snapshot ($o_1, o_2, ..., o_t$), then we pick the replica from the corresponding orbit $o_t$.

DP can also be applied to select orbits. For each time slot $t$ and each orbit $o$, we first calculate the best satellite replica with the lowest query cost, denoted as $v_{o,t}$. Then, we determine the best orbit we can choose in every time slot. Here, choosing orbit $o$ in time slot $t$ means deploying the replica $v_{o,t}$. In other words, we quantify the quality of the orbit based on the best replica in the orbit. The orbit selection DP utilizes $g(t, o)$ to represent the minimum cost for time $1,2,...,t$ and orbit $o$ must be chosen at time $t$. We enumerate the chosen orbit in $t - 1$ to minimize $g(t, o)$.

Given the orbit selection $o_1, o_2, ..., o_t$, and the current replica set $\{S_1, S_2, ..., S_T\}$, we require that only satellite from orbit $o_t$ can be added to $S_t$ when constructing the nearby set. We do not allow deletion and replacement operation to reduce computation cost. With the above changes, we arrive at the multi-time orbit-based local search (MTOLS) algorithm as shown in Algorithm \ref{alg:mtols}.

\begin{algorithm}[H]
\caption{Multi-time Orbit-based Local Search (MTOLS)}
\label{alg:mtols}
\begin{algorithmic}[1]
\Require Time slot number $T$, Origin server set $V_{origin}$, Replica candidate set ${V_{replica}}$, Query cost function $QC$, Replication cost function $RC$, Storage cost function $SC$, Max iteration number $M$
\State $S_t = V_{origin}, t=1,2,..,T$
\For{$m = 1$ to ${M}$} \Comment{iteratively update $S_t$}
    \State $g(0, \emptyset) = 0$ \Comment{orbit selection DP}
    \For{$t = 1$ to $T$}
        \For {each $o_t$ at time $t$}
            \For {each $\hat{o}_{t-1}$ at time $t-1$}
                \State $\hat{c} = QC(t, o_t) + SC(t, o_t) + RC(t, \hat{o}_{t-1}, o_t) + g(t-1, \hat{o}_{t-1})$
                \State $g(t, o_t)=min\{\hat{c}, g(t, o_t)\}$
            \EndFor
        \EndFor
    \EndFor
    \State Get $o_t, t=1,2,..,T$ according to $g$
    \State Update $S_t$ similarly with Algorithm \ref{alg:mtls} but only adding one replica from $o_t$ is allowed to generate $S^{nearby}_t$
\EndFor
\Ensure $S_t, t=1,2,..,T$
\end{algorithmic}
\end{algorithm}




Each iteration in MTOLS consists of one orbit selection DP and one replica selection DP. Suppose $P$ is the orbit number, and $Q$ is the satellite number in each orbit. The orbit selection DP takes $O(TP^2)$, and the replica selection DP takes $O(TQ^2)$. Therefore, the total time complexity of MTOLS is $O(MT(P^2 + Q^2))$, where $M$ is the maximum iteration number. In contrast, the time complexity of MTLS is $O(MTk^2N^2)$, where $N=PQ$. MTOLS is faster than MTLS because $P^2 + Q^2 \ll P^2Q^2 = N^2 < k^2N^2$. The inequality $P^2 + Q^2 < P^2Q^2$ holds when $P > 2$ and $Q > 2$. As an example, we have $P=72$ and $Q=22$ for Starlink phase I satellites. MTOLS will be $\frac{22^2\cdot72^2}{22^2 + 72^2} \cdot k^2 \approx 442 \cdot k^2$ times faster than MTLS theoretically when applied to Starlink phase I.

The core idea of MTOLS is to first determine the orbit, then determine the satellite. It can be seen as an effective pruning of MTLS. MTLS treats every satellite equally. However, in MTOLS, we first identify on which orbits satellites with lower query cost or replication cost are likely to exist, and rule out the orbits that are not promising. After we prune the orbits, we carry out a fine-grained search in these orbits to obtain the best solution.  Our experiments show that MTOLS can achieve the performance of MTLS quite well, while obtaining a speedup of hundreds of times.

\section{Evaluation}

We conduct comprehensive experiments to evaluate the performance of our methods in comparison to baselines and uncover key features of satellite networks. Experiments include trace-driven simulations and prototype implementation. 




\subsection{Experiment Setup}

\noindent \textbf{Satellite Constellations:} Satellite constellations can be roughly categorized into 3 types by the orbit height: LEO, MEO, and GEO satellites. In this paper, we want to examine all of these satellite constellations. We select StarLink phase I for LEO, O3b for MEO, and ViaSat for GEO. StarLink phase I consists of 1,584 satellites in 72 orbits at 550km height \cite{starlinkphase1}. O3b, owned by SES S.A., is a MEO constellation of 20 satellites operating at 8,062km \cite{huang2020recent}. ViaSat consists of 4 geostationary satellites \cite{viasat}. All three constellations provide Internet access and have real-world applications. We enable inter-satellite connections for Starlink, with each satellite connected to 4 neighboring satellites in its orbit and adjacent orbits \cite{chaudhry2021laser}. O3b and ViaSat do not have inter-satellite links.




\noindent \textbf{Gateways:}  As illustrated in Figure \ref{figure:user-satellite-typical-link}, the client connects to satellites, which then connect to gateways, also known as ground stations. The gateways dispatch client requests to the Internet. 166 gateways for Starlink were collected from a website (starlink.sx) and used in our experiment.


\noindent \textbf{Trace Datasets:} We conduct trace-driven simulation using three real traces: MAWI \cite{mawi}, Wikipedia \cite{song2020learning}, and CAIDA \cite{caida}. MAWI and CAIDA are packet traces collected in monitored links in Japan and the US, respectively, and are filtered to only include web requests by checking the protocol and port used in packets. The Wikipedia dataset is collected from a west-coast machine in the US which serves multi-media content for Wikipedia. This dataset does not contain geographical information of visitors, thus we randomly assign the requests to each US state based on population distribution and exclude Alaska and Hawaii, as they are out of the service region of Starlink or ViaSat.  We only keep the top 10 most visited content and 1-hour trace for all of these three datasets. Besides, we also use some synthetic datasets, which will be introduced in specific experiments.

\noindent \textbf{Sampled Real Latency:} We consider 3 types of metrics: hop count, ideal latency and sampled real latency. Hop count is straightforward: satellite-to-user, satellite-to-gateway, and every inter-satellite link are all counted as 1 hop each. Ideal latency is calculated using physical distance and the speed of transmission. Additionally, real latency measurements were conducted in a Starlink network. We acquire the latency of the connection between the user and a gateway by using traceroute. \highlight{Latency measurements are taken every 1 second for 1 day, and we randomly sample latency from measured satellite-to-ground links in our experiments.}

\noindent \textbf{Replication Cost Ratio:} Query cost can be measured by latency or hop count. Besides, proper settings for replication cost and storage cost are also necessary. The transfer cost between nodes $v_1$ and $v_2$ is set to be $\alpha$ times of the query cost between these nodes, where $\alpha > 1$. The intuition here is that replicating between nearby satellites is cheaper and more convenient than between remote satellites. To avoid frequent replica position changes, $\alpha > 1$ is used, with $\alpha = 50$ as the default value in experiments. A value of $\alpha = 1$ would result in replicating the origin server to a nearby satellite for each user, yielding the same cost as directly querying from the origin server. A larger $\alpha$ requires a careful selection of replica positions to serve a large number of users for a long time.


\noindent \textbf{Storage Cost Ratio:} We define the smallest query cost across all time stamps to be $c_{qmin}$. It will be 1 or several milliseconds if hop count or latency is used as metrics, respectively. The storage cost is set to $\beta \cdot c_{qmin}$ for gateways and $\gamma \cdot c_{qmin}$ for satellites, where $\gamma > \beta$. We refer to $\beta$ and $\gamma$ as storage ratios. The difference between $\gamma$ and $\beta$ reflects the cost difference between terrestrial and satellite storage. Usually satellite storage is much more expensive than terrestrial storage. In our evaluation, $\beta=1$ and $\gamma=10$ are used based on the data from \cite{lai2021cooperatively}, \highlight{which shows that storing and querying 1 GB of data costs \$0.1 from terrestrial CDN operators and \$1 for satellite network providers.} 

\subsection{Algorithm Evaluation}
\subsubsection{Baseline methods}

In this section, we introduce the baseline methods. In general, our baselines contain not only heuristic algorithms for UFL (uncapacitated facility location problem) \cite{arya2001local, jain2003greedy}, but also algorithms designed for a satellite network \cite{lai2021cooperatively, pfandzelter2021edge}.

\noindent \textbf{Algorithms for UFL:} Our problem can be reduced to a UFL problem if only one time slot is considered. Therefore, we can employ an algorithm of UFL and apply it at each time slot independently.


We evaluate the following three popular UFL algorithms: \highlight{First, a naive greedy algorithm tries to add one new facility in every iteration. The location of the new facility is chosen greedily to minimize the total cost. The algorithm stops when the total cost cannot be reduced.} Second, a smarter greedy algorithm proposed in \cite{jain2003greedy} uses an average cost instead of the total cost as the greedy metrics. When selecting new facility location, it considers (i) the average opening cost and transport cost to unassigned users, and (ii) the reduction in transport cost for assigned users. This algorithm has been shown to be a 1.61-approximation of the UFL problem and will be referred to as the "1.61x greedy" in our experiments. Third, a local search algorithm \cite{arya2001local} has an approximation factor of 3 for the UFL problem, as proved in \cite{gupta2008simpler}. Some other algorithms based on linear programming rounding reach lower approximation factors \cite{li20131,byrka2007optimal}. We do not include these methods because we have thousands of satellites and running linear programming on these many facilities will take too long.

\noindent \textbf{StarFront:} \cite{lai2021cooperatively} proposes StarFront to set up replicas on satellites and cloud servers to minimize the latency. The algorithm needs a latency threshold for all users and it chooses replicas for every user to satisfy the latency threshold. If multiple replicas satisfy the threshold, the one with the smallest sum of replication cost and storage cost will be chosen. Our metrics include not only latency but also hop count. Thus we extend this algorithm to support hop count. We set an appropriate threshold by minimizing the total cost.

\noindent \textbf{Periodic Cache Handoff (PCH):} \cite{pfandzelter2021edge} proposes a heuristic to manage satellite cache. It will place the cache near the end users first. Then the cache will be replicated to the next satellite on the same orbit periodically. There is also inter-orbit cache replication with a lower frequency. We refer to this method as periodic cache handoff (PCH). 

\subsubsection{Results of Different Metrics}
\label{sec:basic-overview}

\begin{figure}[t]
\centering
\includegraphics[width=0.475\textwidth]{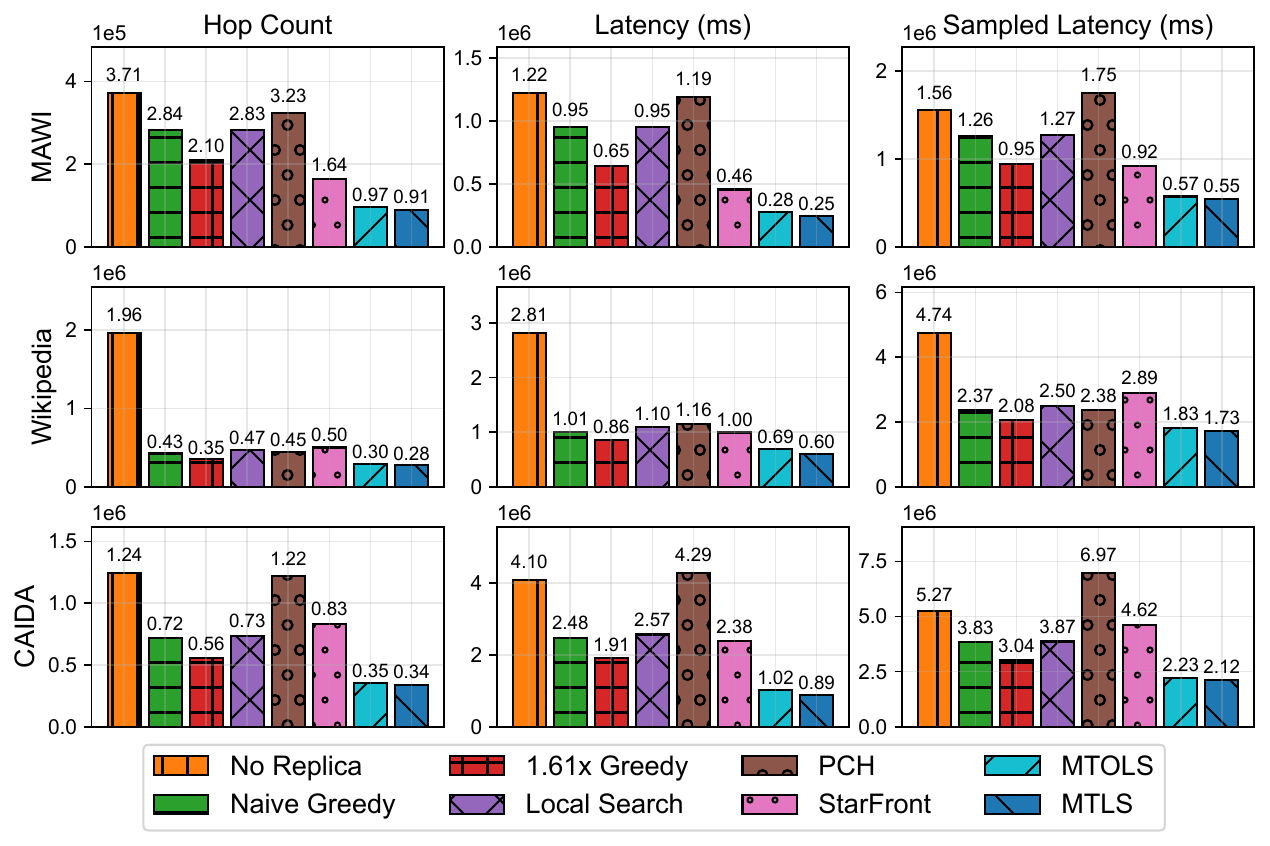}
\caption{Evaluation of different methods based on 3 real traces (MAWI, Wikipedia, CAIDA) is presented in the figure. The total cost of various metrics is displayed in columns.
}
\label{figure:basic-overview}
\end{figure}



Our results, shown in Figure \ref{figure:basic-overview}, demonstrate the effectiveness of our proposed methods, MTLS and MTOLS. The experiments include LEO satellites and gateways in the network structure. 
 
Among the evaluated algorithms, our methods consistently achieved the lowest total cost across 3 trace datasets and 3 metric types. MTLS outperformed all other algorithms with a reduction in total cost ranging from 16.91\% to 53.26\% compared to the strongest baseline. MTOLS also showed favorable performance compared to all baselines. 

\begin{figure}[t]
\centering
\includegraphics[width=0.475\textwidth]{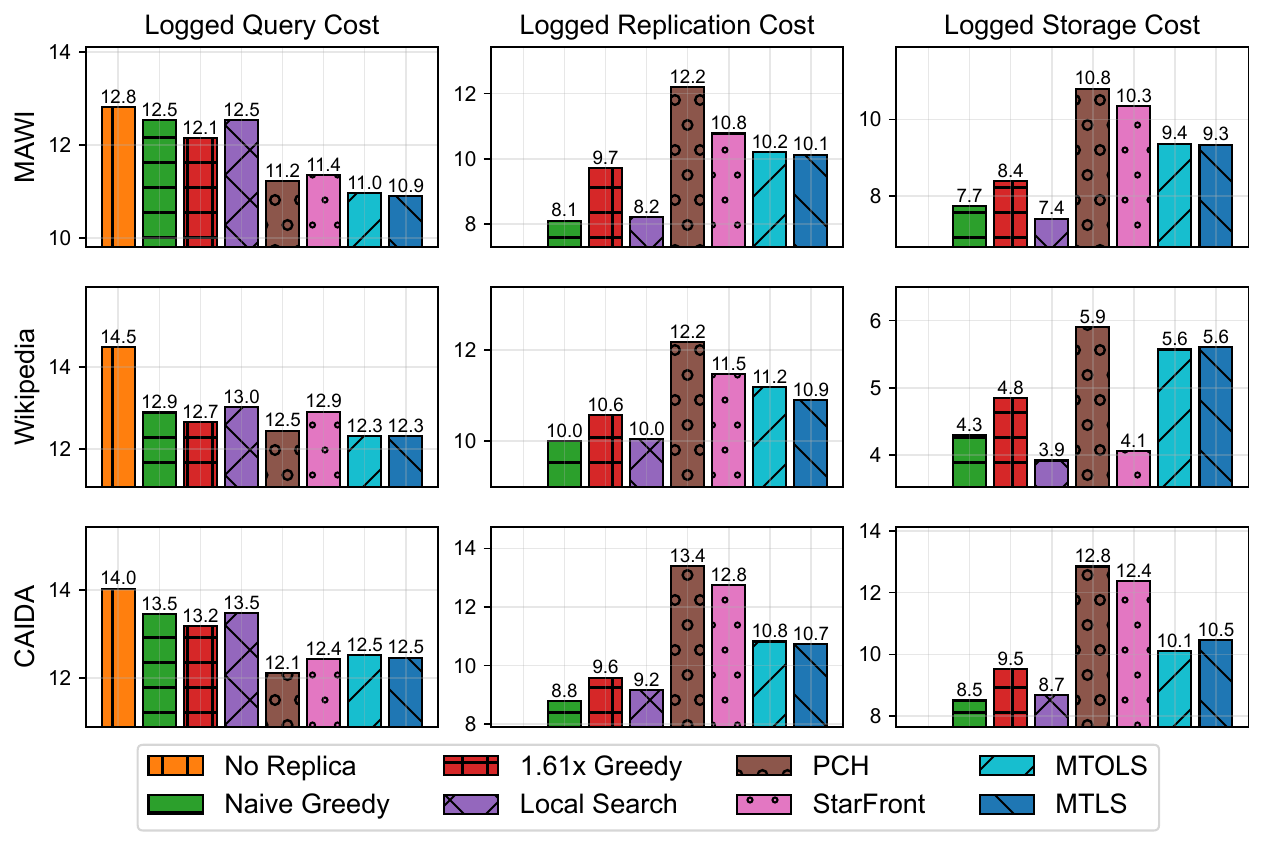}
\caption{This figure shows the breakdown of the total cost: Query cost, replication cost and storage cost. The metric is hop count.
 }
\label{figure:basic-breakdown-metrics_hop_count}
\end{figure}

The total cost of each method is decomposed into query cost, replication cost, and storage cost, as shown in Figure \ref{figure:basic-breakdown-metrics_hop_count}. It reveals that UFL algorithms typically have lower replication and storage costs, but higher query costs. This is due to the UFL algorithms' focus on demand within each time slot without considering future demand. They are designed to optimize for static networks and cannot handle dynamic network and dynamic user demands. 

In contrast, PCH and StarFront are designed for satellite network, but each has notable drawbacks. As noted in \cite{pfandzelter2021edge}, the intra-orbit replication of PCH takes place every 4.3 minutes for all replicas, leading to high replication cost. This is evident in Figure \ref{figure:basic-breakdown-metrics_hop_count}. Such aggressive replication produces unnecessary traffic because: (1) Satellites passing through hot regions don't need to replicate their content to the next satellite—it can be used later. (2) Satellites in less popular regions could discard content instead of transferring cache. On the other hand, StarFront doesn't allow changes to replicas once established. To achieve optimal latency, it sets up numerous replicas, thereby incurring significant replication and storage costs. Unlike PCH and StarFront, our methods balance all costs effectively. We also find MTLS has lower replication costs than MTOLS. The reason may be that MTOLS does not have replacement operations but MTLS has it when generating the nearby set.

\subsubsection{Oracle v.s. Prediction of Demand}
\label{sec:oracle-vs-pred}
\begin{figure}[t]
\centering
\includegraphics[width=0.475\textwidth]{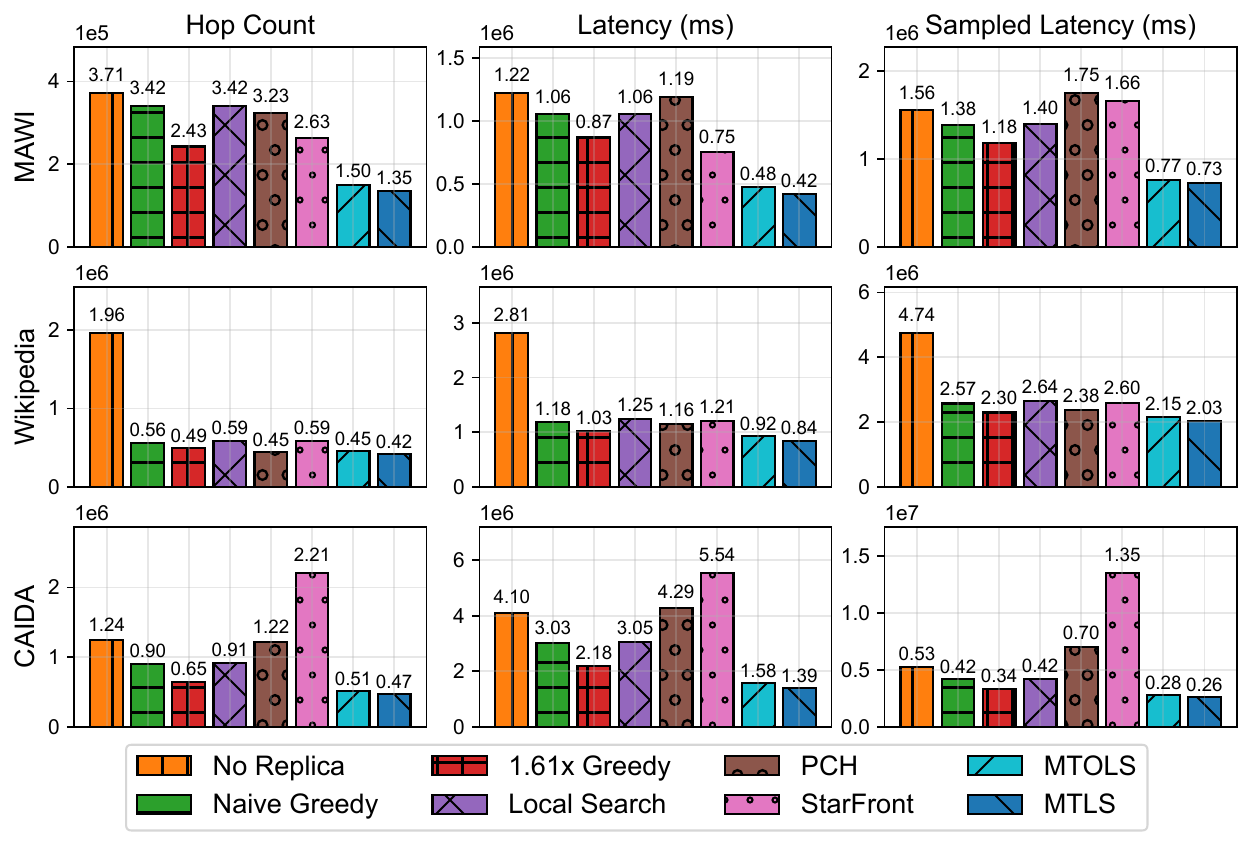}
\caption{Evaluation of methods without oracle information of demand. Predicted demand based on average historical values is used for each method.
}
\label{figure:pred-overview}
\end{figure}



In Section \ref{sec:basic-overview}, the experiments used oracle information of demand to select replicas, but this information is not always available in real-world scenarios. We conduct additional experiments without oracle information, but using averaged historical demand data over the past 5 minutes to predict future demand. This prediction is used as input for all algorithms except PCH, which is rule-based and does not require predictions. Figure \ref{figure:pred-overview} shows that the gap between PCH and other methods narrows, but our methods continue to outperform the other algorithms. This demonstrates their robustness. To further optimize the system, our methods can be combined with more advanced prediction methods designed for CDNs, such as the stack distance model \cite{mattson1970evaluation,almasi2002calculating,bjornsson2013dynamic}, heuristics \cite{hu2019proactive}, the Markov \cite{berger2017adaptsize}, time series \cite{hassine2017arma},  and machine learning models \cite{song2020learning}. Given the previous requests to different content from different regions, these methods are employed to decide whether we should keep the content for each region, or predict the expected demand of the content in different regions. Then our methods can be used to determine an efficient placement in satellite networks.

\subsubsection{Computation time}
\label{sec:computation-time}

\begin{table}[t]
\caption{Computation time of different methods on the MAWI dataset.}
\begin{tabular}{cc}
\hline
\textbf{Method}       & \textbf{Computation Time (sec)} \\ \hline
No Replica            & 0.0           \\
Naive Greedy          & 20.9          \\
1.61x Greedy & 30.1          \\
Local Search          & 33.9          \\
PCH                   & 0.3           \\
StarFront             & 13.8          \\
MTOLS                 & 495.3         \\
MTLS                  & 98,576.3      \\ \hline
\end{tabular}
\label{table:basic-time}
\end{table}

We show the computation time on the MAWI dataset in Table \ref{table:basic-time}. All methods are evaluated on the same machine with Intel Xeon CPU E5-2690 v3 CPU and 220 GB memory size. Although the CPU has 24 vcores, we limit the program to use only up to 1 vcore for comparison. We find that MTOLS can have 200x speedup over MTLS by incorporating orbit information. It can be further accelerated using parallel execution. We can optimize the placement of different content in parallel and make different processes to handle different subsets in the DP phase in parallel.

\subsubsection{Combining Different Satellite Constellations}
\label{sec:leo-meo-geo}

\begin{figure}[tp]
\centering
\includegraphics[width=0.45\textwidth]{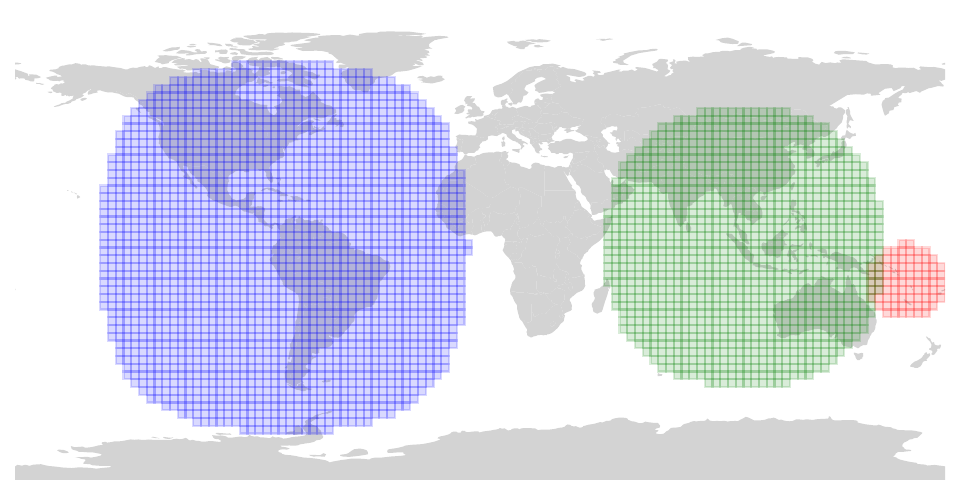}
\caption{The coverage map of one Starlink (LEO), O3b (MEO), and ViaSat (GEO) satellite, colored in red, green, and blue, respectively. We set the minimum elevation angle to 10 degree for all satellites for fair comparison.}
\label{figure:leo-meo-geo-coverage}
\end{figure}

\begin{table}[]
\caption{The results of integrating MEO and GEO in the network, with MTLS determining satellite selection based on different storage cost factors. The results include query cost, replication cost, and storage cost in the last three columns, with the time ratio of MEO and GEO usage shown in the "MEO" and "GEO" columns.}
\label{tab:meo-geo}
\begin{tabular}{clccccc}
\hline
\multicolumn{1}{l}{\textbf{$\gamma_{meo}$}} & \textbf{$\gamma_{geo}$} & \multicolumn{1}{l}{\textbf{MEO}} & \multicolumn{1}{l}{\textbf{GEO}} & \multicolumn{1}{l}{\textbf{Qry.}} & \multicolumn{1}{l}{\textbf{Rep.}} & \multicolumn{1}{l}{\textbf{Sto.}} \\ \hline
0                              & 10         & 100.0\%                          & 0.0\%                            & 18,060                              & 300                                   & 0                                    \\
0.5                            & 10         & 100.0\%                          & 0.0\%                            & 18,060                              & 300                                   & 18                                   \\
1                              & 10         & 38.9\%                           & 61.1\%                           & 18,010                              & 150                                   & 234                                  \\
2                              & 10         & 19.4\%                           & 80.6\%                           & 18,000                              & 100                                   & 304                                  \\
3                              & 10         & 0.0\%                            & 100.0\%                          & 18,000                              & 50                                    & 360                                  \\
5                              & 10         & 0.0\%                            & 100.0\%                          & 18,000                              & 50                                    & 360                                  \\
10                             & 10         & 0.0\%                            & 100.0\%                          & 18,000                              & 50                                    & 360                                  \\ \hline
\end{tabular}
\end{table}


\begin{figure*}[t]
\centering
\includegraphics[width=0.85\textwidth]{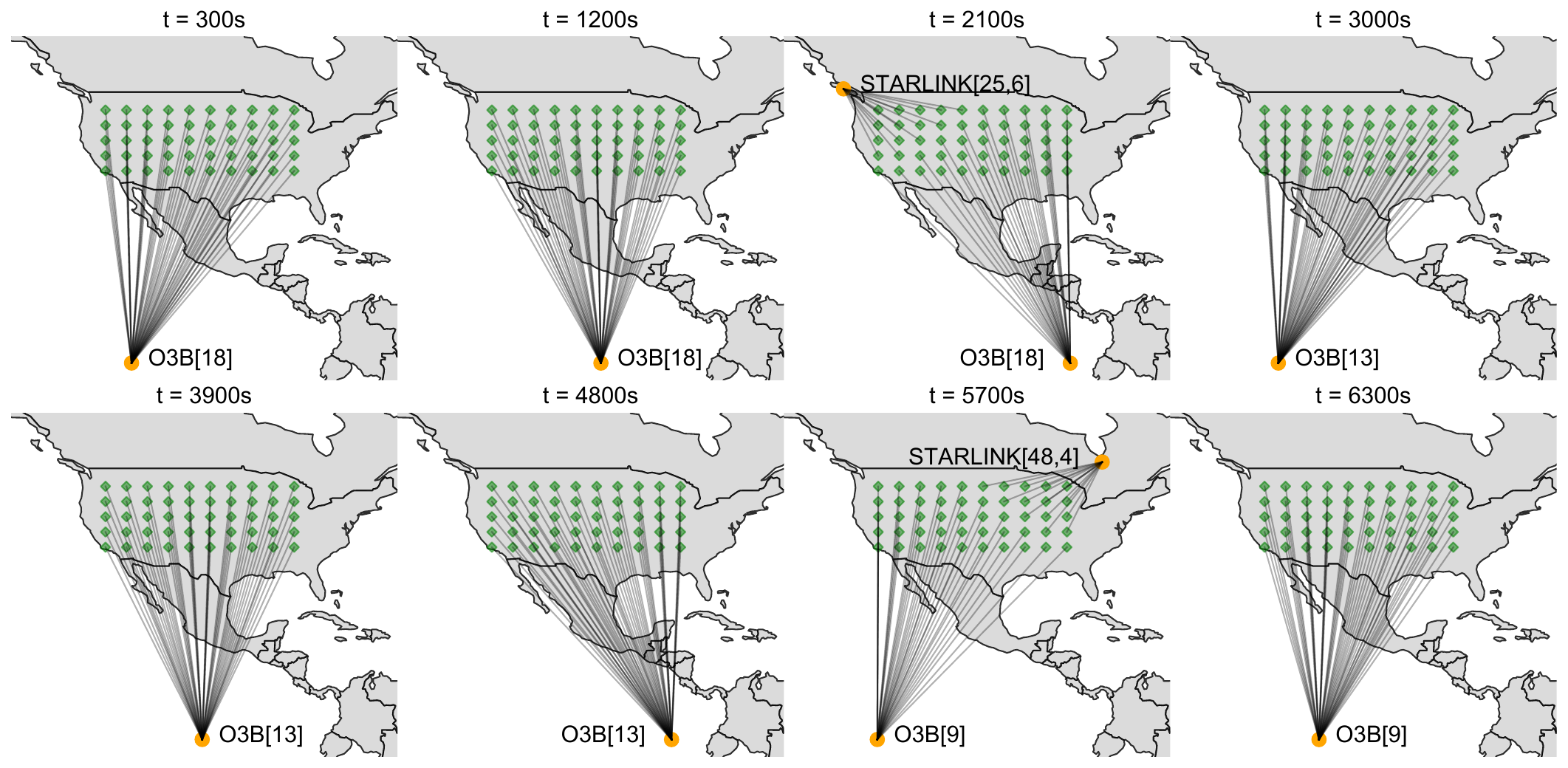}
\caption{
We put LEO and MEO satellites in the same network, and let MTLS decides which kind of satellites to use. We set different storage cost factor $\gamma$ for LEO and MEO: $\gamma_{leo}=1$, and $\gamma_{meo}=50$. In this scenario, LEO satellites and MEO satellites can be cooperatively deployed as replicas to cover all of the demand.
} 
\label{figure:leo-meo-coop}
\end{figure*}

\begin{table}[]
\caption{
The results of integrating LEO and MEO in the network, with MTLS determining satellite selection. We set different storage cost factor $\gamma$ for LEO and MEO: $\gamma_{leo}=1$, and $\gamma_{meo}=25$. The results include query cost, replication cost, and storage cost in the last three columns, with the time ratio of LEO and MEO usage shown in the "LEO" and "MEO" columns.
}
\label{tab:leo-meo}
\begin{tabular}{cccccc}
\hline
\textbf{Region} & \textbf{LEO} & \textbf{MEO} & \textbf{Qry.} & \textbf{Rep.} & \textbf{Sto.} \\ \hline
1x2             & 100.0\%      & 0.0\%        & 24,000         & 1,200             & 48               \\
2x4             & 39.6\%       & 60.4\%       & 24,000         & 600               & 744              \\
3x6             & 2.1\%        & 97.9\%       & 24,000         & 300               & 1,176            \\
4x8             & 2.1\%        & 97.9\%       & 24,016         & 300               & 1,176            \\
5x10            & 2.1\%        & 97.9\%       & 24,080         & 450               & 1,176            \\ \hline
\end{tabular}
\end{table}

We compare the performance of three types of satellite networks: LEO (Starlink), MEO (O3b), and GEO (ViaSat). Figure \ref{figure:leo-meo-geo-coverage} visualizes the coverage map for these three types of satellites. For a fair comparison, we set a unified minimum elevation angle (10\textdegree) for all satellites. We also remove all gateways from the network structure. The experiment is conducted using a simulated dataset of requests from a 5x10 grid region in the US over 4 hours. We assume each region has the same volume of requests every 5 minutes.



The results indicate that the choice of satellites depends on storage cost factors and the chosen metric: either hop count or latency. When storage cost factors for LEO, MEO, and GEO are identical, the preferred network is GEO for hop count metric because of its large coverage range, and LEO for latency metric due to its close distance to the ground. 


Evaluation using the hop count metric with different storage cost factors for MEO and GEO shows that a reduction in MEO's storage cost factor leads to more frequent use of MEO, as shown in Table \ref{tab:meo-geo}. The coverage areas of MEO and GEO satellites are similar, however, MEO satellites move slowly and require content replication to the next satellite when moving away from the US. The likelihood of using MEO increases when the reduction in storage cost compensates for the increase in replication cost.

MEO has a larger coverage area than LEO. We examine the impact of the geographical distribution of demand on satellite selection by limiting demand to 1x2, 2x4, 3x6, 4x8, and 5x10 out of the 5x10 grid region. Storage cost factors of LEO and MEO are set at 1 and 25, respectively. Table \ref{tab:leo-meo} shows that LEO is preferred for serving small regions of demand, while MEO is preferred for serving larger regions. Meanwhile, if the storage cost of LEO is low enough, both LEO and MEO are used as shown in Figure \ref{figure:leo-meo-coop}.

\subsubsection{Effects of Geographical Locations}
\label{sec:geographical}



\begin{figure}[tp]
\centering
\includegraphics[width=0.45\textwidth]{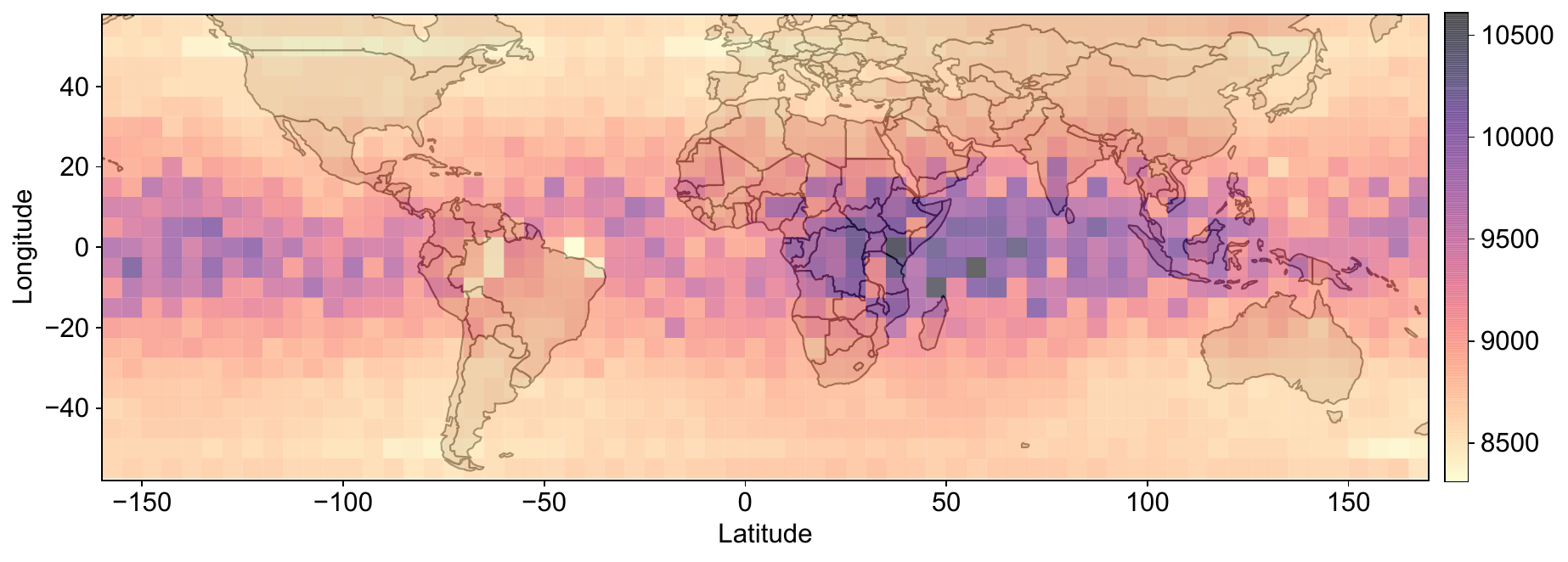}
\caption{One client is set at different locations and MTLS is used to calculate total cost for serving it. }
\label{figure:across_lat_lon}
\end{figure}



We study the total cost for serving a single client placed at varying locations using MTLS with only Starlink satellites and Starlink gateways. Figure \ref{figure:across_lat_lon} shows that higher latitudes have less total cost than lower latitudes, which is likely due to the increased visibility of satellites \cite{mcdowell2020low}. Servicing countries at the same latitude in South America costs less than in Africa, likely due to fewer gateways in Africa.


\subsubsection{Effects of gateways}
\label{sec:gateway}
\begin{figure}[tp]
\centering
\includegraphics[width=0.5\textwidth]{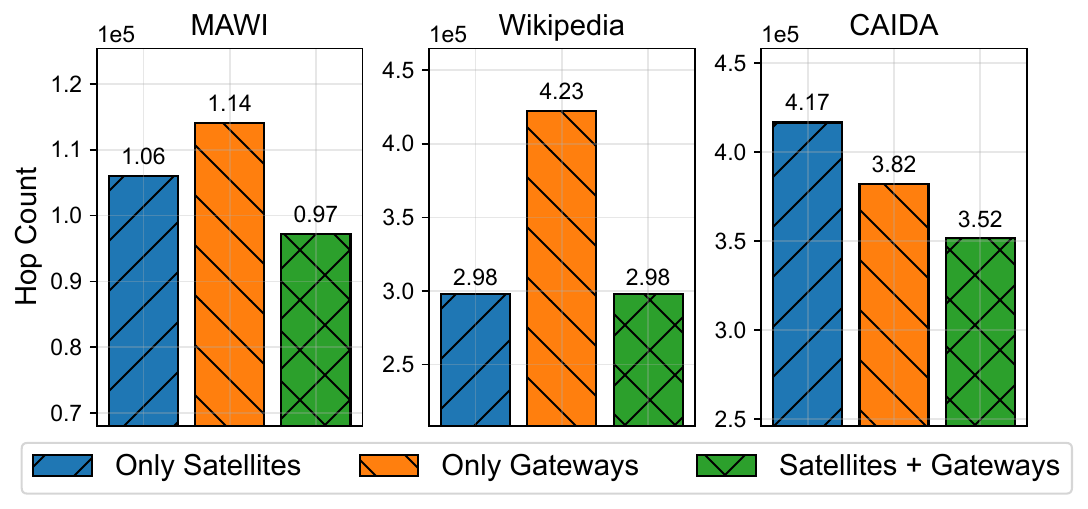}
\caption{Results of limiting MTLS to choose replicas only from gateways or only from satellites. Hop count is used as a metric.}
\label{figure:gateway-satellite-coop}
\end{figure}

\begin{figure*}[t]
\centering
\includegraphics[width=0.95\textwidth]{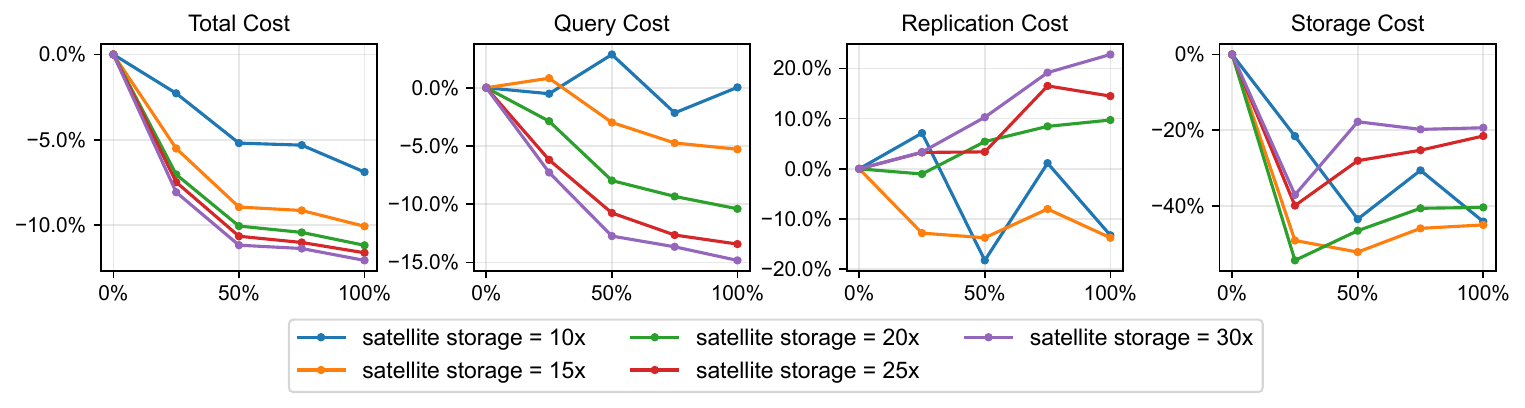}
\caption{Results of adding new gateways near clients. In the experiment, we place new gateways near 0\%, 25\%, 50\%, 75\%, and 100\% of the clients. Satellite storage is set to be 10, 15, 20, 25, and 30 times of gatway storage. CAIDA is used as the assessed dataset and hop count is employed as the metric.}
\label{figure:add-gateway}
\end{figure*}


To examine the impact of gateways, we maintain the same network structure but force the algorithm to deploy replicas only on gateways or only on satellites. 
Figure \ref{figure:gateway-satellite-coop} shows that using both gateways and satellites is preferred for the MAWI and CAIDA datasets. On the Wikipedia dataset, utilizing only satellites is sufficient.  The reason is that the requests of the Wikipedia dataset are concentrated in the US and LEO satellites alone are sufficient to cover its demands. One limitation of a satellite replica is its limited coverage area. This disadvantage is mitigated in the case of concentrated requests, \eg, on the Wikipedia dataset. The tradeoff between gateways and satellites is that satellites can provide less latency, less hop count but larger storage cost and moving coverage area, while gateways can provide smaller storage cost, static coverage area but more latency and hop count. 


We further evaluate the impact of adding gateways to a satellite network using the CAIDA dataset. We use hop count as the metric. Figure \ref{figure:add-gateway} demonstrates that adding gateways reduces the overall cost especially when the cost of satellite storage is much higher than gateway storage. With moderate satellite storage costs (\eg, 10$\times$ or 15$\times$), satellites are used as intermediate transfer points: The replica is first transmitted to a satellite, then the satellite travels to a specific region, and finally the replica is unloaded from the satellite to a gateway in that region, thereby reducing the replication cost. \highlight{\cite{dowhuszko2020leo} proposes a similar solution to migrate hot cached content in 5G edge nodes by transmitting it to a LEO satellite. The content is then multicasted to different 5G nodes as the LEO satellite moves to another region. But different from \cite{dowhuszko2020leo}, our algorithm is more general and can adapt to different scenarios.} The above strategy is less preferred when the satellite storage becomes more expensive (\eg, 20x, 25x, 30x). In this case, our algorithm prioritizes deploying replicas on more gateways, which leads to increased replication cost but decreased query cost.

\subsubsection{Video Delivery}
\label{sec:video-delivery}
We test the performance of different methods for video delivery. We create video demand by randomly generating video chunk downloading requests and assigning them to different states of the US according to population distribution. We exclude Alaska and Hawaii. 



The download time of each chunk is calculated as the sum of propagation delay and transmission delay, where propagation delay is determined by distance and transmission delay is determined by chunk size and throughput. We assume 20 Gbps for terrestrial links and 10 Gbps for satellite links.

In addition to redirecting the user to the closest server, we use round robin and weighted round robin for redirection. We take $n=3$ for these 2 strategies. For weighted round robin, the 3 closest replicas will get $\frac{4}{7}$, $\frac{2}{7}$, and $\frac{1}{7}$ traffic, respectively. 

Table \ref{tab:simulated-video} summarizes our results. MTOLS and MTLS usually have better QoE than the baselines because our methods can generate relatively well-distributed replicas, which results in a more balanced traffic distribution in the network. Although the local search has lowest traffic but has lower QoE. The traffic of our methods is lower than all baselines except local search when "closest" or weighted round robin is used as the traffic strategy. For most algorithms, a weighted round robin is better than "closest" and round robin.

\begin{table*}[]
\caption{The results of the simulated video delivery. The Mean QoE denotes the average quality of experience at each time slot, and the Traffic GB indicates the cumulative transmitted traffic in all links. If one piece of content is transmitted through multiple links, the traffic will be counted for multiple times. 
}
\label{tab:simulated-video}
\begin{tabular}{cccccccc}
\hline
\multirow{2}{*}{\textbf{Method}} & \multirow{2}{*}{\textbf{Replica Num}} & \multicolumn{2}{c}{\textbf{Closest}} & \multicolumn{2}{c}{\textbf{Round Robin}} & \multicolumn{2}{c}{\textbf{Weighted Round Robin}} \\
                                 &                                       & Mean QoE        & Traffic GB         & Mean QoE          & Traffic GB           & Mean QoE               & Traffic GB               \\ \hline
No Replica                       & 1.0                                   & 1.70            & 963,159            & 1.70              & 963,159              & 1.70                   & 963,159                  \\
Naive Greedy                     & 6.5                                   & 5.40            & 308,808            & 4.19              & 405,530              & 5.12                   & 404,049                  \\
1.61x Greedy            & 11.4                                  & 6.25            & 320,389            & 5.40              & 425,899              & 6.28                   & 412,831                  \\
Local Search                     & 5.3                                   & 4.82            & \textbf{293,368}   & 3.35              & \textbf{376,532}     & 4.29                   & \textbf{387,299}         \\
PCH                              & 18.5                                  & 6.41            & 304,255            & 6.33              & 416,780              & 6.89                   & 397,025                  \\
StarFront                        & 26.0                                  & 3.18            & 318,911            & 4.75              & 476,172              & 4.51                   & 447,326                  \\
MTOLS                            & 16.9                                  & \textbf{6.63}   & 295,152            & \textbf{6.59}     & 416,093              & \textbf{7.12}          & 389,084                  \\
MTLS                             & 18.3                                  & \textbf{6.63}   & 295,313            & 6.54              & 418,526              & 7.08                   & 389,732                  \\ \hline
\end{tabular}
\end{table*}

\subsection{System Evaluation}

\begin{figure}[tp]
\centering
\includegraphics[width=0.4\textwidth]{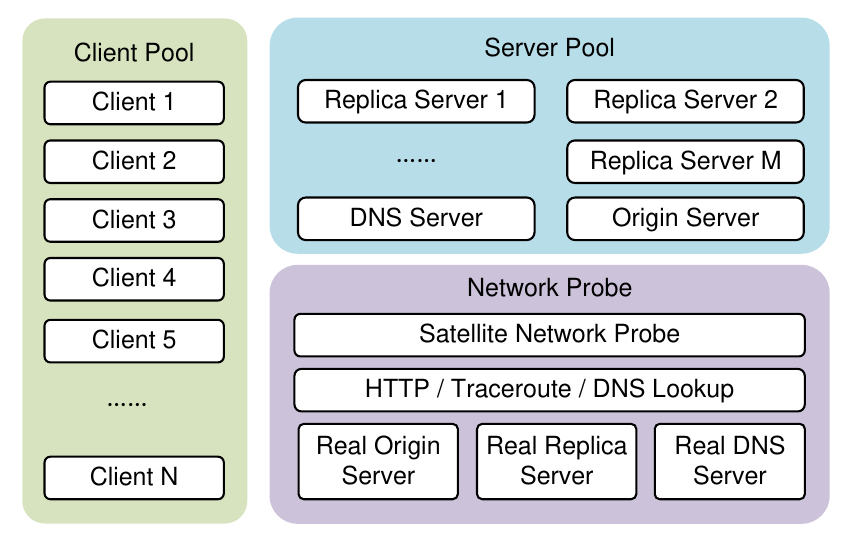}
\caption{Prototype System Implementation}
\label{figure:real-system-design}
\end{figure}

We evaluate our system through a prototype implementation. As we do not have control access to the satellite network (\eg, replicating content to real satellites), we emulate a satellite network using network traces and let user requests go through network links according to the traces. Figure \ref{figure:real-system-design} shows our system. It consists of 3 parts: the client pool, the server pool, and the network probe. The client pool and the server pool are process pools. User clients, replica servers, the DNS server, and the origin server run as separate processes. Network performance metrics like bandwidth and latency are measured using real-world data from the Starlink networks, alongside traceroute for satellite-to-gateway latency. An authoritative DNS server is set up to direct user queries to the appropriate replica servers. This setup aims to provide a realistic evaluation of the system by closely mimicking actual satellite network conditions.

In our experiments, the network probe is conducted within a Starlink RV satellite network in the United States. The probe finds a bandwidth of 4.90±1.12 Mbps to a Web server and 11.76±3.11 Mbps to the closest Akamai server, with a latency of 129.79±92.2 ms to the nearest Starlink ground station.

\subsubsection{Web Browsing Results}

\begin{figure*}[t]
\centering
\includegraphics[width=1.0\textwidth]{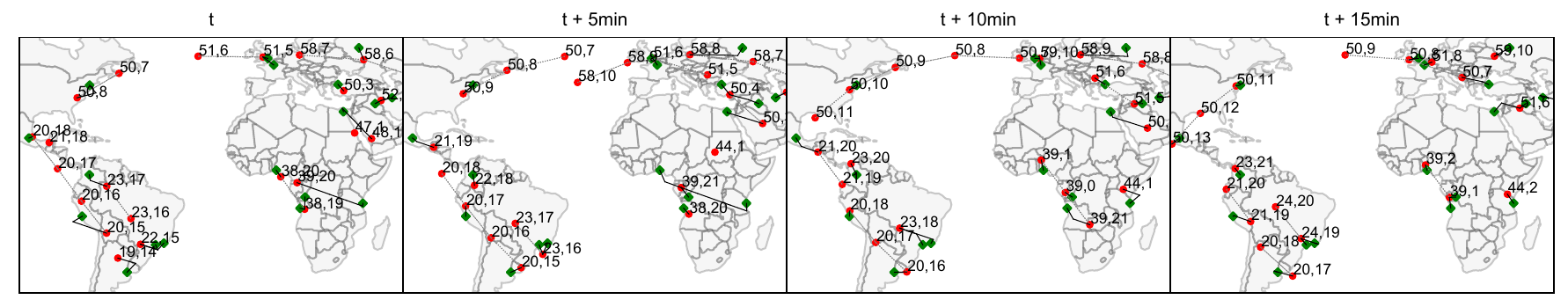}
\caption{\highlight{Four snapshots from the solution of the web browsing scenario where MTLS generates satellite replicas for the top 50 most populated global cities. Red dots denote satellite replicas, while green diamonds denote the targeted cities. Solid lines depict city-satellite connections, and inter-satellite connections are marked by dotted lines. Satellites are marked by $i, j$, where $i$ is the orbit index and $j$ is the index in that orbit.}}
\label{figure:trace_analysis}
\end{figure*}

In order to optimize the web browsing experience, real-world CDNs are utilized to host static web content, such as images, CSS files, HTML documents, and fonts on replicas to minimize the page load time experienced by end users. Our study assesses the proposed system's performance for web browsing by letting clients download a static item of size 16KB from the servers in our prototype system. We generate simulated requests from the top 50 populated cities on Earth. The replica placement is determined by applying various algorithms to the dataset. The resulting replicas are deployed in the server pool, and clients in the user pool are instructed to request from the servers. The download time is recorded and is presented in Table \ref{tab:real-web}. MTLS has the smallest download time. In addition, MTOLS exhibits a similar download time to PCH, but with a more efficient deployment of 25.0 replicas compared to PCH's 34.9 replicas. In every approach, a portion of request time is allocated to DNS queries. The fixed satellite trajectories allow for pre-fetching future server information, mitigating the necessity for frequent DNS queries. Experimental data indicates that, with DNS queried every 10 minutes, it takes up less than 15\% of the total request time across all methods.

As a case study, we present four representative snapshots of satellite replicas chosen by MTLS in Figure \ref{figure:trace_analysis}. The solution exhibits two patterns: (1) Satellite orbits traversing more targeted regions are prioritized. Orbit $50$, for instance, traverses Mexico City, New York, London and Paris. Satellite replicas $50,8$ on orbit $50$ are maintained across all intervals, which serves New York at $t$, London at $t$ + $15$min, and provisions replica $50, 9$ at at $t$ + $5$min. It effectively reduces both replication and query costs. Similar strategies are employed in South American for orbits $20$ and $24$. (2) Regions that are not reachable through a single efficient orbit are serviced by individual satellites with cache periodically transferred between satellites on the same orbit. Dar es Salaam, an African city, employs this strategy where cache is transitioned from $44,1$ to $44,2$ from $t$ + $10$min to $t$ + $15$min. This strategy aligns with the human-designed method in \cite{pfandzelter2021edge}, but our method detects these patterns automatically. Besides, \cite{pfandzelter2021edge} usually incurs high replication cost as the handoff is designed to be periodical. Our method seeks to reduce the cost by reusing the same satellite without handing off the cache.

\begin{table}[h]
\caption{Performance of web browsing median download time. "Num" is the average number of deployed replicas in each time slot. "DNS Per." is the proportion of DNS time in the request time. StarFront[MTLS] and StarFront[MTOLS] indicate aligned replica numbers with MTLS and MTOLS.}
\label{tab:real-web}
\centering
\begin{tabular}{cccc}
\hline

\textbf{Method}                & \textbf{Num} & \textbf{DNS Per.} & \textbf{Time (ms)} \\ \hline
No Replica                     & 1.0          & 0.0\%    & 2089.5             \\
Naive Greedy                   & 8.3          & 6.3\%    & 177.0              \\
A 1.61x Approx Greedy          & 10.9         & 7.4\%    & 151.1              \\
Local Search                   & 4.4          & 5.8\%    & 184.9              \\
Periodic Cache Handoff         & 34.9         & 12.2\%   & 108.7              \\
StarFront                       & 43.0         & 11.1\%   & 126.0              \\
StarFront {[}MTOLS{]} & 26.0         & 9.9\%    & 136.9              \\
StarFront {[}MTLS{]}  & 38.0         & 10.7\%   & 129.9              \\
MTOLS                          & 25.0         & 12.0\%   & 109.2              \\
MTLS                           & 37.5         & 13.2\%   & \textbf{96.5}               \\ \hline

\end{tabular}
\end{table}

\subsubsection{Video Streaming Results}

\begin{table}[h]
\caption{
Performance of video streaming. "Num" is the average number of deployed replicas in each time slot. StarFront[Aligned] means we align the number of replicas set by StarFront to be the same with MTLS and MTOLS.
}
\label{tab:real-video-streaming}
\centering
\begin{tabular}{cccc}
\hline
\textbf{Method}                & \textbf{Total Cost} & \textbf{Mean QoE} & \textbf{Num} \\ \hline
No Replica                     & 30628.5             & 1.56              & 1.0                  \\
Naive Greedy                   & 2612.1              & 8.00              & 6.3                  \\
1.61x Greedy          & 2431.3              & 8.74              & 8.2                  \\
Local Search                   & 3058.2              & 7.23              & 4.8                  \\
PCH                            & 2714.8              & 8.93              & 17.6                 \\
StarFront                      & 3634.6              & 8.70              & 29.0                 \\
StarFront {[}Aligned{]} & 4386.2              & 6.84              & 11.0                 \\
MTOLS                          & 2403.5              & 9.00              & 10.4                 \\
MTLS                           & \textbf{2281.0}     & \textbf{9.15}     & 10.8                 \\ \hline
\end{tabular}
\end{table}

Furthermore, we also evaluate the performance of our system for video streaming. We adopt similar settings to  Section \ref{sec:video-delivery} but make the following two modifications to the settings: (1) we limit the capacity of each server to 96 Mbps, instead of applying a capacity limit on connections, and (2) we randomize the start time for each user's video stream. This allows us to reduce the total number of concurrent requests so that we can run the experiments on our machines.  Table \ref{tab:real-video-streaming} summarizes our results. Our methods -- MTOLS and MTLS out-perform the other algorithms in terms of total cost and QoE. While MTOLS and MTLS have similar query costs (2055 and 2025), they differ in terms of replication cost (343 and 250) and QoE (9.00 and 9.15). Our hypothesis is that optimizing the replication cost in the network can lead to a more balanced distribution of traffic, thereby improving the overall QoE.

\section{Related Work}
\label{sec:related}


\highlight{
\noindent \textbf{CDN optimization:} Much research has focused on CDN server placement optimization, targeting efficient replica or content positioning given constraints like bandwidth, latency, and replica count. The problem is modeled as facility location \cite{sung2013efficient}, K-median \cite{placement01}, and K-center \cite{jin2016toward}. The problem is NP-hard and algorithms vary in performance metrics, constraints, target networks, and effectiveness \cite{replica-survey}. Most algorithms suit static networks, but they don't apply to LEO or MEO satellite networks. \cite{lai2021cooperatively} and \cite{pfandzelter2021edge} consider replica server or cache placement in LEO network but our methods go beyond these work by (i) leveraging orbit information for scalability, (ii) using DP for efficient placement, (iii) supporting both webpages and videos, and (iv) supporting various satellite types (LEO, MEO, GEO, and combinations).



\noindent \textbf{Satellite Network:} Recently there is a surge of interest in LEO satellite networks as the launch and hardware cost rapidly decreases. For example, \cite{Handley-hotnets18} develops a simulator for Starlink and shows that it can provide lower latency than a terrestrial optical fiber network. \cite{StarPerf} develops StarPerf to simulate mega-constellation and uses it to guide constellation design. Studies have been made to enhance LEO networks, such as adding low-cost, download-only ground stations \cite{L2D2}, improving delay and throughput via multiple links \cite{OrbitCasts}, and developing strategies for real-time traffic relay node selection and flow allocation \cite{SpaceRTC}. Other works suggest replica placement strategies on satellites \cite{lai2021cooperatively}. As shown in our evaluation, our approach significantly outperforms \cite{lai2021cooperatively}.
}

\section{Conclusion}
\label{ssec:conclusion}

In this paper, we develop novel replica servers for satellite networks. We leverage the deterministic satellite movement trajectories along with the user demands to move content close to the users when needed. Using both synthetic and real measurements from StarLink, we evaluate our approach across diverse network scenarios. Our results show our placement yields 16.91\% to 53.26\% reduction in the total cost while maintaining low query latency and high video QoE. Using prototype implementation and experiments, we further show the feasibility and effectiveness of our approach. 

\printbibliography


\end{document}